# Unveiling the Regulatory Factors for Phase Transitions in Zeolitic Imidazolate Frameworks: A High-Throughout Calculations and Data Mining

Zuhao Shi[1], Bin Liu[1], Yuanzheng Yue[1,3], Arramel Arramel[5], Neng Li[1,2,4*]

[1]State Key Laboratory of Silicate Materials for Architectures, Wuhan University of Technology, Wuhan 430070, China

[2]State Key Laboratory of Advanced Technology for Float Glass Technology, CNBM Bengbu Design & Research Institute for Glass Industry Co., Ltd, Bengbu 233010, China

[3]Department of Chemistry and Bioscience, Aalborg University, DK-9220 Aalborg, Denmark

[4]State Center for International Cooperation on Designer Low-Carbon & Environmental Materials (CDLCEM), School of Materials Science and Engineering, Zhengzhou University, Zhengzhou 450001, Henan, China

[5]Nano Center Indonesia, Jalan Raya PUSPIPTEK, South Tangerang, Banten 15314, Indonesia

*Correspondence: lineng@whut.edu.cn

**Abstract:** Recently, there have been significant advancements in the study of Metal-Organic Frameworks (MOFs), particularly in the discovery of glassy states in zeolitic imidazolate frameworks (ZIFs), a subset of MOFs. However, the correlation between the glass-forming ability (GFA) of MOFs and their structural characteristics remains poorly understood. To address this gap, we analyze over 20,000 MOFs from the QMOF database, comparing geometric and electronic order parameters between ZIFs and other MOFs. Our findings reveal a set of order parameters that effectively capture the thermal stability and GFA of MOFs, acting as structural genes to predict glass formation. Through molecular dynamics simulations on representative ZIF structures, we validate the reliability of these order parameters, particularly the total bond-order density (TBOD) and the largest cavity diameter (LCD), for screening potential melt-quenched MOFs. Furthermore, we predict that external pressure and electric fields can control the TBOD of MOFs, expanding our understanding of their thermal stability. This work identifies structural genes associated with the melt stability of MOFs, and contributes valuable insights into the prediction and understanding of MOF glass-forming ability.

**Keywords**: *MOF Glasses; Reactive Molecular Dynamic; DFT Simulation; Data Learning.*

**Introduction**

Research on amorphous Zeolite Imidazolate Frameworks (ZIFs) has emerged as a prominent area of investigation within the realm of organic metal frameworks in recent years.[1-3] In contrast to their crystalline counterparts, amorphous ZIFs exhibit intriguing research potential, such as superior gas adsorption properties [4] and enhanced mechanical strength.[5,6] Notably, glassy ZIFs have significantly advanced our understanding of glassy materials owing to their distinctive organic-inorganic hybrid chemical composition.[3] Currently, the predominant methods employed to synthesize glassy-state ZIFs revolve primarily around the melt-quenching approach.[7] For ZIFs exhibiting strong glass-forming ability, such as ZIF-4 and ZIF-62, glassy ZIFs structures preparation can be achieved via the melt-quench method.[8-11] However, for ZIFs with limited glass-forming ability like ZIF-8, chemical modification techniques utilizing polar solvents or water can be employed to significantly reduce the melting point below the decomposition temperature, thereby enabling the formation of stable melts that can subsequently be cooled to obtain glassy samples.[12,13] It is anticipated that future endeavors will primarily focus on mitigating thermal decomposition and enhancing melt stability during the preparation of ZIF glasses. Despite the potential demonstrated by in-situ thermal processing techniques for fabricating glass structures,[14] their widespread implementation is currently constrained by higher costs. In comparison, the melt-quenching method has attained a commendable level of proficiency in the production of large-scale glass and film materials.[15,16]

Although new glass-state Zeolitic Imidazolate Frameworks (ZIFs) are being continuously explored, they remain relatively scarce compared to the extensive family of Metal-Organic Frameworks (MOFs). For most MOFs, the major challenge for glass formation is the low thermal stability. At high temperatures, MOFs undergo thermal decomposition instead of transforming into stable melts. Healy et al. have provided a systematic exposition on the thermal stability of MOFs, highlighting that the relative strengths of nodes and organic linkers are key influencing factors.[2] They claimed ZIFs consist of weaker metal nodes but stronger organic ligands compared to MOFs. As a consequence, organic ligands in ZIFs can withstand higher temperatures without undergoing decomposition, demonstrating an empirical criterion consistent with the melting mechanism of crystals. Recent simulation studies and chemical spectroscopy research have further corroborated the micro-mechanism of ZIFs' melting.[17,18] The exchange of functional groups at the central node under high temperatures, resulting in the occurrence of missing ligands.[19] This short-range exchange interaction between metal nodes and organic ligands is regarded as a characteristic reaction of molten ZIFs.[3] Melting induces damage to the crystal structure, and localized damage in the original grains leads to a reduction in the enthalpy of fusion and a subsequent decrease in the melting point. When the melting point falls below the thermal decomposition temperature, the MOF structure can achieve stable melting. The weaker interaction between metal nodes and organic ligands renders the bonds more susceptible to thermal cleavage than the intramolecular bonds in organic ligands. Upon lowering the temperature, some of the missing ligands and over-coordinated nodes are retained, contributing to the unique optical properties exhibited by ZIFs.[20,21]

The simulation works on short-exchange interaction provides a microscopic perspective on the melting behavior of ZIFs, but only offers a preliminary insight into the molten structure of ZIFs. To establish a comprehensive quantitative structure-activity relationship (QSAR) for the short-range interactions inherent in the ZIFs melting process, we must further consider the topological structure information of MOF structure. For instance, apart from the influence of different metal nodes and chemical bonds, the diverse topologies of MOFs also significantly impact the process. In their unmodified state, the sod topology exhibits a greater propensity for thermal decomposition rather than melting, in contrast to the cag structure.[7,10] This distinction is attributed to the proportion and distribution of pores within the frameworks. Nevertheless, a dearth of systematic investigations exists concerning the influence of distinct types of order parameters in different MOFs on the melting behavior of ZIFs.

In this work, we employ a comprehensive approach integrating data analysis and atomistic simulation methods to examine the melting process in MOFs and endeavors to establish a QSAR to elucidate MOF thermal stability. Taking advantage of the Quantum MOF (QMOF) database,[22] which encompasses the electronic structure information of MOFs, we explore the distinctions between ZIFs and other MOFs through the analysis of various structural order parameters. Subsequently, employing reactive molecular dynamic simulations, we investigate the heating mechanisms of sixteen different ZIFs. By combining geometric and electronic structure order parameters with ZIFs' thermal behavior, we propose order parameters capable of characterizing the melting propensity of MOFs, thus offering theoretical guidance for the screening of MOF structures with heightened glass-forming ability. Finally, combined with Density Functional Theory (DFT) calculations, we validate several potential strategies for enhancing the melt stability of ZIFs.

**Results**

In this investigation, we embark on a comprehensive analysis aiming to elucidate the intercorrelation among various order parameters. In the pursuit of comprehending the nuanced disparities existing within the realm of MOFs and ZIFs, a meticulous examination is undertaken to discern and elucidate the intricate variances pertaining to both geometric and electronic structural order parameters. This endeavor is carried out with the utmost precision, focusing on several pivotal structural characteristics that bear significant importance in this context. The utilization of a heatmap, as depicted in **Fig 1(a)**, provides a visual representation of the interdependence among the diverse structural parameters under consideration. By amalgamating the correlation analysis of these parameters with fundamental principles rooted in the domain of physical chemistry, we are able to discern significant structural descriptors that exhibit independence from one another.

Regarding the geometric structural parameters, our findings unveil a positive correlation between the Pore Limiting Diameter (PLD) and the Largest Cavity Diameter (LCD) with three normalized energy values, specifically $E_{total}$/Volume, $E_{vdw}$/Volume, and $E_{elec}$/Volume. These normalized energy values signify diverse atomic interactions transpiring within a given unit volume. Specifically, $E_{elec}$/Volume serves as an indicator of the strength of short-range forces, more specifically the bonding strength of chemical bonds. Conversely, $E_{vdw}$/Volume is determined by calculating the ratio of van der Waals interaction energy to volume, employing DFT-D3 calculations, thereby providing insights into the magnitude of long-range forces within a unit volume.[23] It is worth noting that higher values of $E_{vdw}$/Volume correspond to stronger van der Waals interactions.

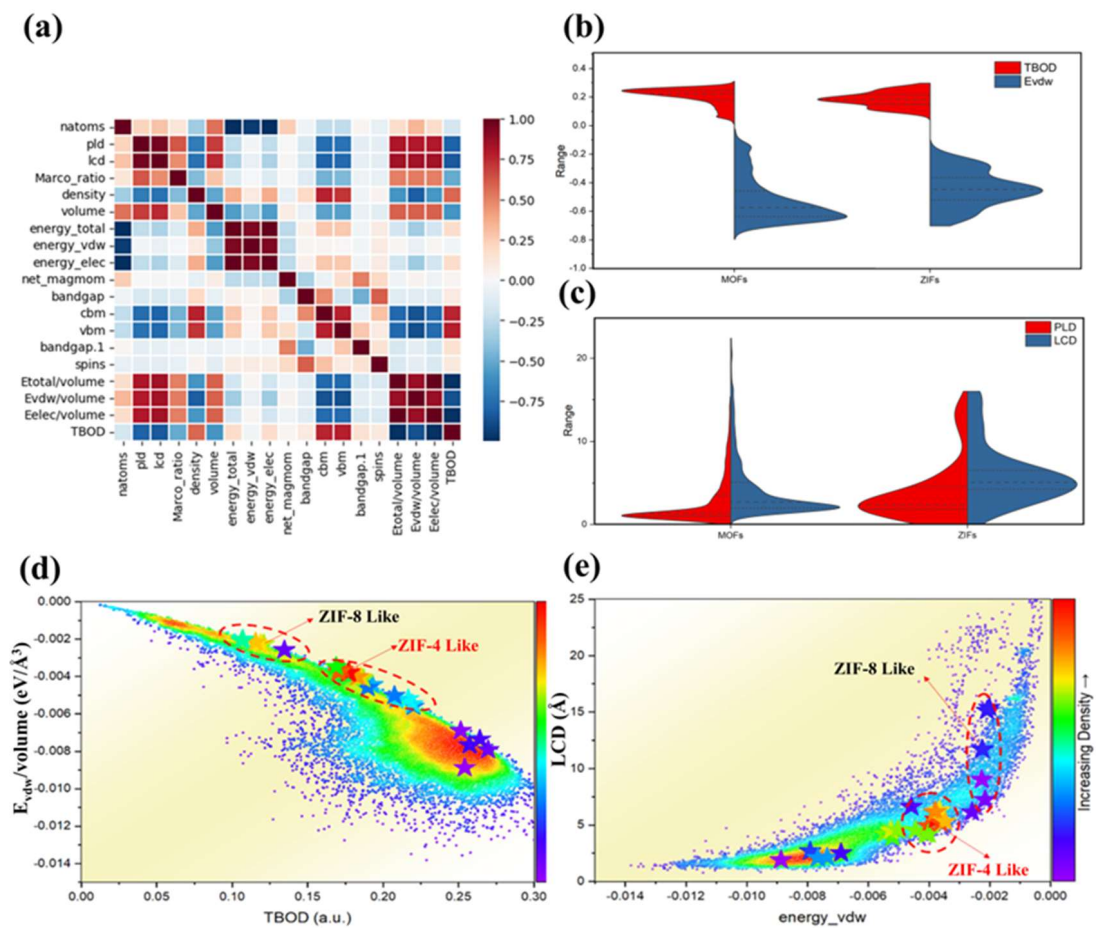

**Fig 1. Analysis of the correlation and distribution characteristics between the various order parameters of the MOFs.** (a) Heatmap of correlations between all structural parameters. The color bar on the right indicates the correlation ranging from negative to positive, where blue represents a negative correlation between the vertical and horizontal sequential parameters, and red indicates a positive correlation. Violin plots of MOFs and ZIFs depicting the distribution of (b) TBOD and $E_{vdw}$/Volume, and (c) PLD and LCD; (d) Kernel density distribution plot of MOFs on TBOD and $E_{vdw}$/Volume; (e) Kernel density distribution plot of MOFs on $E_{vdw}$/Volume and LCD.

As elucidated in the Methods section, TBOD, a composite parameter evaluated through the DDEC method for bond strengths, assumes a pivotal role in characterizing the internal cohesion strength of amorphous oxides.[24-26] By considering the treatment of atomic charges, TBOD is posited as a quantity that mirrors the intensity of short-range interactions within distinct MOFs. Remarkably, our findings reveal a conspicuous negative correlation between PLD, LCD, and TBOD. This indicates that MOFs endowed with larger pore sizes exhibit diminished TBOD, thereby reflecting the intrinsic association between porosity and cohesion strength. Notably, additional parameters, such as the conduction band minimum (cbm) and valence band maximum (vbm), offer more pronounced correlations, as they encapsulate band characteristics. However, these parameters were excluded as effective variables for subsequent investigations due to the unclear physical mechanism elucidating how the band gap influences crystal melting and the acknowledged underestimation of the band gap by the PBE functional. Likewise, density, despite being a fundamental physical property, was not further scrutinized owing to the considerable impact of different elements in MOFs on density, which is accounted for by TBOD and reduced energy values. In summary, based on the correlation analysis conducted among the variables, we ultimately identified four effective parameters for subsequent studies: PLD and LCD, which collectively capture the overall crystal geometry, and TBOD and $E_{vdw}$/Volume, which respectively signify the strength of short-range and medium-range interactions within the crystal.

The majority of meltable MOFs fall within the subclass of ZIFs, as elucidated in the Introduction section. Consequently, it becomes imperative to investigate the disparities in the effective structural parameter distributions of ZIFs and MOFs. **Fig 1(b)** presents the distribution violin plots illustrating the variations in TBOD (Total Bond Order Density) and $E_{vdw}$/Volume (reduced van der Waals energy per unit volume) for both MOFs and ZIFs, with dashed lines denoting the quartile and median lines, respectively. In terms of TBOD distribution, ZIFs exhibit lower values compared to MOFs, indicative of diminished cohesive strength in short-range interactions within ZIFs. Regarding mid-range interaction strength, ZIFs display a lower reduced van der Waals energy when compared to MOFs, implying a lower average molecular weight of organic ligands. When hydrogen bonding and the influence of chemical composition are not taken into account, it is commonly observed that diminished magnitudes of van der Waals forces are correlated with reduced melting points. By amalgamating the differences in TBOD and van der Waals interaction strength distributions, it can be inferred that, in contrast to MOFs, ZIFs exhibit weaker bonding interactions in both short-range and mid-range structures. **Fig 1(c)** showcases the distribution disparities of geometric parameters PLD and LCD between MOFs and ZIFs. ZIFs demonstrate higher PLD and LCD values compared to MOFs, indicating the presence of larger pores within ZIFs. Taking into consideration the intrinsically transient nature of short-range exchange interactions unfolding at metal nodes in the course of ZIFs' melting process, it is posited that the presence of larger pores within these structures may bestow an augmented realm for the facilitation of organic functional group exchanges.

To more intuitively reflect the distribution characteristics of effective structural parameters of MOFs and ZIFs, the kernel density distribution plots of effective structural parameters are drawn in **Fig 1(d)** and **(e)**, using the color bar to represent the probability distribution. A few ZIFs structures are marked with pentagrams. Furthermore, we highlighted two typical ZIFs with the same topology: ZIF-8 Like and ZIF-4 Like. Since the QMOF database does not contain ZIFs containing mixed organic ligand compositions, such as ZIF-62 and ZIF-76, we adopt a topology-based approach to our discussion. ZIF-8 Like represents ZIFs based on imidazolate ligands and *sod* topology, while ZIF-4 Like is similar to ZIF-4 and has *cag* topology. It can be seen that compared to MOFs, ZIF-8 Like and ZIF-4 Like have higher $E_{vdw}$/Volume and lower TBOD. Compared to ZIF-8 Like, ZIF-4 Like has a larger TBOD and smaller pore characteristics (LCD). Combining experimental studies on the glass-forming ability of the two typical ZIFs, ZIF-4 Like has a stronger glass-forming ability than ZIF-8 Like, which usually decomposes rather than exists as a stable melt at high temperatures. Based on the above discussions, we believe that the glass-forming ability of ZIFs can be described by TBOD.

Based on our preceding discourse, it is posited that certain electronic structural parameters TBOD and Evdw/Volume, along with geometric structural parameters LCD and PLD, are pertinent factors influencing the melting behavior of MOFs. Through the utilization of the meticulously selected effective structural parameters, one can actively participate in qualitative dialogues aimed at elucidating the discrepancies that distinguish MOFs from ZIFs. Nonetheless, it is worth noting that a quantitative discourse pertaining to precise MOFs is currently absent. Moreover, a comprehensive quantitative analysis, tailored specifically to individual MOFs, remains to be accomplished, leaving a gap in the current body of research. In experimental studies, debates surrounding the criteria for evaluating the glass formation propensity of MOFs are still ongoing, with the ratio of glass transition temperature ($T_g$) to melting point ($T_m$) commonly employed as a benchmark. Nevertheless, certain aspects necessitate clarification. Firstly, the scope of experimental reports on diverse ZIFs is presently limited to only a handful of variants. Moreover, both $T_g$ and $T_m$ are somewhat influenced by experimental conditions, such as pressure and thermal history. Consequently, establishing a quantitative structure-activity relationship (QSAR) between MOF structure and glass formation propensity based solely on existing experimental investigations proves challenging. Consequently, in order to fill this gap, we undertook simulations on a defined scale to establish a correlation between TBOD and the high-temperature behavior of ZIFs. The thermal stability during temperature elevation was simulated employing molecular dynamics with a reactive force field, accounting for the need for extended relaxation time during the heating process. This methodology has been validated as efficacious in elucidating the thermal melting behavior of ZIFs in prior studies.[27-29]

Simulation can provide more structural information on short range (< 5 Å) and medium range (5-20 Å) intuitively, compared with experimental methods. Herein we discuss the thermodynamic behavior based on changes in radial distribution function, through the heating process of sixteen ZIFs from 300 to 1000 K. Meanwhile, we attempt to establish a QSAR for ZIFs based on the four effective structural order parameters selected in the previous section. **Fig S1** illustrates the alterations in density observed within ZIFs as the heating process ensues, concurrently displaying the corresponding values of the four pertinent effective order parameters. Overall, the thermal behavior of different ZIFs can be divided into two categories, corresponding to the black and red font descriptions. ZIFs shown in **Fig S1(a)-(i)** exhibit high thermal stability within the range of 300-1000K, characterized by maintaining stable density over an extended temperature window or a gradual decline observed in correspondence with increasing temperature. We refer to this type of ZIFs as thermally stable ZIFs. On the other hand, the density changes of seven ZIFs shown in **Fig S1(j)-(p)** indicate an increase in density with temperature in the early stages of heating. We refer to this type of ZIFs as thermally sensitive. Considering the structural optimization and long-term dynamic relaxation process at 300K, internal stress in the structures has been effectively eliminated. This density increase with temperature corresponds to thermal expansion of the structure. It can be seen that thermally stable ZIFs usually have higher TBOD (> 0.20) and van der Waals energy density ($|E_{vdw}| > 0.50$ eV/Å$^3$), as well as lower LCD (<6.00) and PLD (< 4.00). Overall, this indicates a correlation between thermal stability and effective order parameters: stronger short- and medium-range interactions, as well as smaller pore sizes.

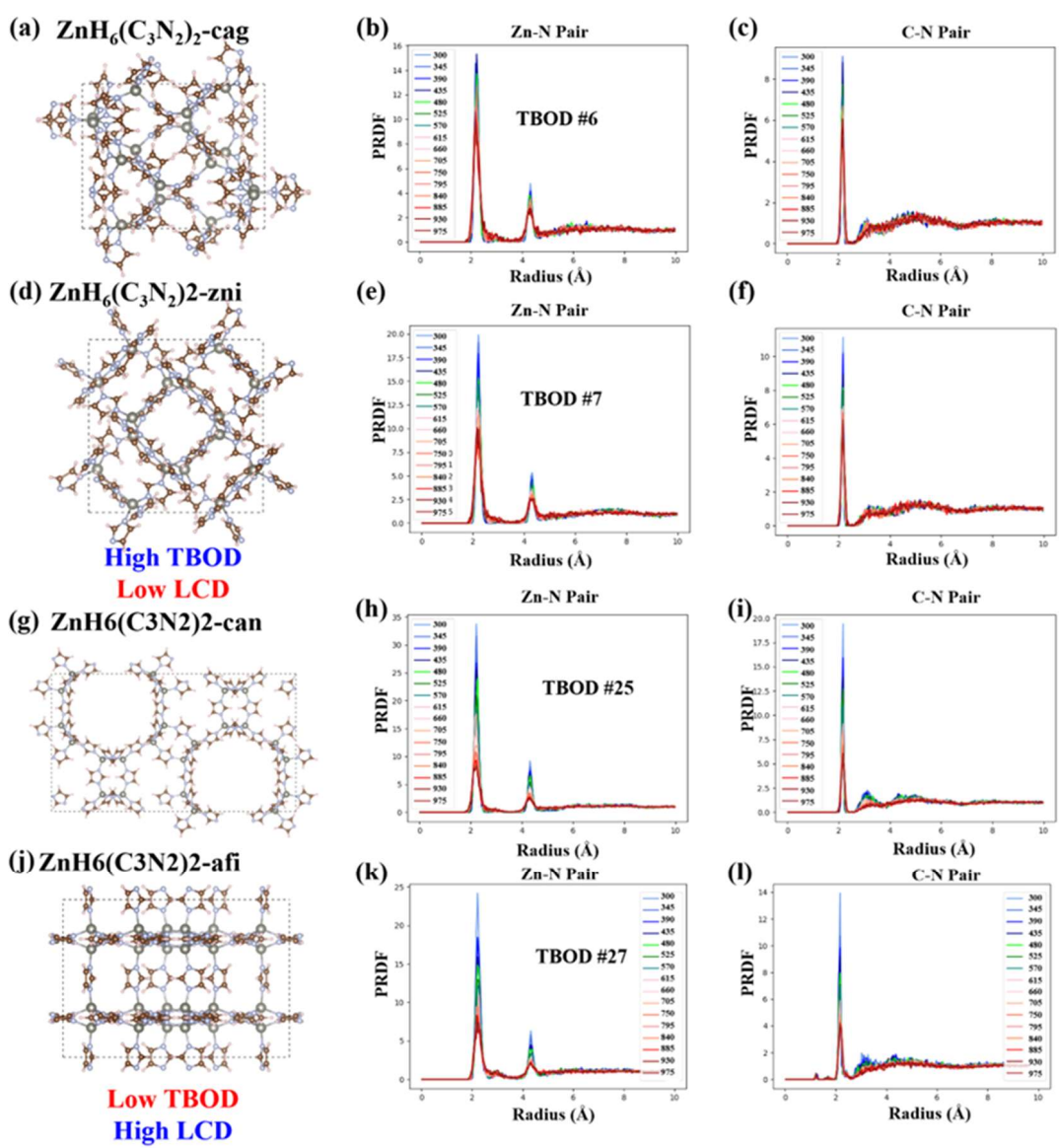

**Fig 2. Variation of radial distribution function (RDF) with temperature for ZIFs with different structural characteristics.** (a) Crystal structure of ZnH6(C3N2)2-cag, and RDFs of (b) Zn-N pair and (c) C-N pair as a function of temperature. (d) Crystal structure of ZnH6(C3N2)2-zni, and RDFs

of (e) Zn-N pair and (f) C-N pair as a function of temperature. (g) Crystal structure of ZnH6(C3N2)2-can, and RDFs of (h) Zn-N pair and (i) C-N pair as a function of temperature. (j) Crystal structure of ZnH6(C3N2)2-afi, and RDFs of (k) Zn-N pair and (l) C-N pair as a function of temperature. The curves transition from blue to red to indicate the temperature range from 300 K to 1000 K.

It is worth noting that the thermal behavior of ZIFs, whether it involves melting or thermal decomposition, is a non-instantaneous process due to the complexity of these crystals. Upon heating, ZIFs initially undergo bond breaking between organic ligand functional groups and metal nodes (intermolecular bond breaking), as well as self-decomposition of organic ligands (intramolecular bond breaking). The ability of ZIFs to exist as stable melts depends primarily on the dominant type of bond breaking. To further investigate the propensity of ZIFs towards stable melting, we conducted an in-depth analysis of the radial distribution function (RDF) for various atomic pairs within the structures. Supporting Information **Figs S2-S17** provide detailed information on the changes in RDFs for sixteen different atomic pairs in ZIFs as a function of temperature. The RDF is a useful tool for characterizing the arrangement of atoms in the short to medium range, where the height and width of the peaks reflect the orderliness of the length distribution for corresponding atomic pairs. In the ideal crystals, atomic pairs engaged in specific interactions display a uniformity in bond lengths, resulting in isolated straight-line peaks. However, amorphous or glass structures with short-range order and long-range disorder typically display a continuous distribution profile in their RDFs. When performing RDF calculations based on molecular dynamics trajectory files, the thermal vibrations of different atomic pairs lead to a certain peak width. A greater degree of thermal vibrations corresponds to wider peaks. Simultaneously, since the number of atom pairs for each type of interaction remains constant (as the total number of atoms is unchanged in the microcanonical ensemble), the areas under each peak remain within a normal range. Consequently, the peak height decreases as the peak width widens. **Fig 2** presents the structures of two representative ZIFs along with their pair radius distribution functions (PRDFs), showcasing the characteristics of variations in different atomic pairs among the sixteen ZIF types. **Table 1** provides information on the TBOD and LCD values for four ZIFs. Notably, the Zn-N pair and C-N pair in the corresponding structures exhibit varying degrees of decrease with increasing temperature. ZnH6(C3N2)$_2$-cag and ZnH6(C3N2)$_2$-zni, characterized by high TBOD and low LCD (**Fig S1(a)** and **(d)**), display a lesser extent of decrease in Zn-N pairs compared to structures with high LCD and low TBOD (ZnH6(C3N2)$_2$-can and ZnH6(C3N2)$_2$-afi). The aforementioned observation provides evidence suggesting that MOFs characterized by robust bonding attributes demonstrate enhanced thermal stability under short-range interactions.

**Table 1. Rankings of four representative ZIFs based on their effective order parameters, TBOD and LCD, among sixteen ZIFs.**

| Species | TBOD (#Rank) | LCD (#Rank) |
| --- | --- | --- |
| ZnH6(C3N2)$_2$-cag | 0.220 (#6/16) | 4.360 (#9/16) |
| ZnH6(C3N2)$_2$-zni | 0.219 (#7/16) | 4.239 (#10/16) |
| ZnH6(C3N2)$_2$-can | 0.119 (#14/16) | 11.729 (#3/16) |
| ZnH6(C3N2)$_2$-afi | 0.107 (#16/16) | 15.416 (#1/16) |

The thermal stability of various structures was quantitatively assessed using specific indicators, building upon the aforementioned discussions. The stability of different atom pairs at distinct temperatures was examined by analyzing the height variation of different peaks in partial radial distribution function (PRDF). Two standard deviations, namely $\sigma_{Zn-Ligands}$ and $\sigma_{Ligands-Ligands}$, were defined to evaluate the short-range structural changes. These standard deviations corresponded to the ratio of peak heights in the first peak of the PRDF at high temperature (1000 K) and low temperature (300 K) respectively. i.e

$$\sigma_{Zn-Ligand} = \frac{First\ Peak\ Height\ of\ Zn - Ligands\ pairs\ (T = 1000\ K)}{First\ Peak\ Height\ of\ Zn - Ligands\ pairs\ (T = 300\ K)}$$

$$\sigma_{Ligands-Ligands} = \frac{First\ Peak\ Height\ of\ Ligands - Ligands\ pairs\ (T = 1000\ K)}{First\ Peak\ Height\ of\ Ligands - Ligands\ pairs\ (T = 300\ K)}$$

the value of $\sigma_{Zn-Ligands}$ served as a measure of the preservation of short intermolecular distances within ZIFs during the heating process. Higher values of $\sigma_{Zn-Ligands}$ indicated a greater retention of the original short-range order, implying a reduced occurrence of Zn-Ligands bond breakage. Similarly, the value of $\sigma_{Ligands-Ligands}$ represented the stability of molecular interactions during the heating process, with larger values suggesting enhanced stability of organic ligands without decomposition. **Fig 3** illustrates the correlation between thermal stability and effective structural parameters of sixteen ZIFs. As depicted in **Fig 3(a)**, an increase in TBOD was observed to result in a more stable Zn-Ligands interaction, indicating enhanced thermal stability of intermolecular ZIFs. Moreover, the stability of $\sigma_{Ligands-Ligands}$ also exhibited an upward trend with increasing TBOD, reflecting improved stability of organic ligand molecules. These findings imply that the thermal stability of both intermolecular and intramolecular ZIFs is augmented with the intensification of structural bonding effects. Furthermore, **Fig 3(b)** demonstrates that ZIFs with larger pore sizes, as indicated by increased LCD values, experienced a decrease in structural thermal stability. This reduction was associated with a higher occurrence of bond cleavage phenomena at elevated temperatures, encompassing both metal-organic ligand connections and ligand intramolecular bonds.

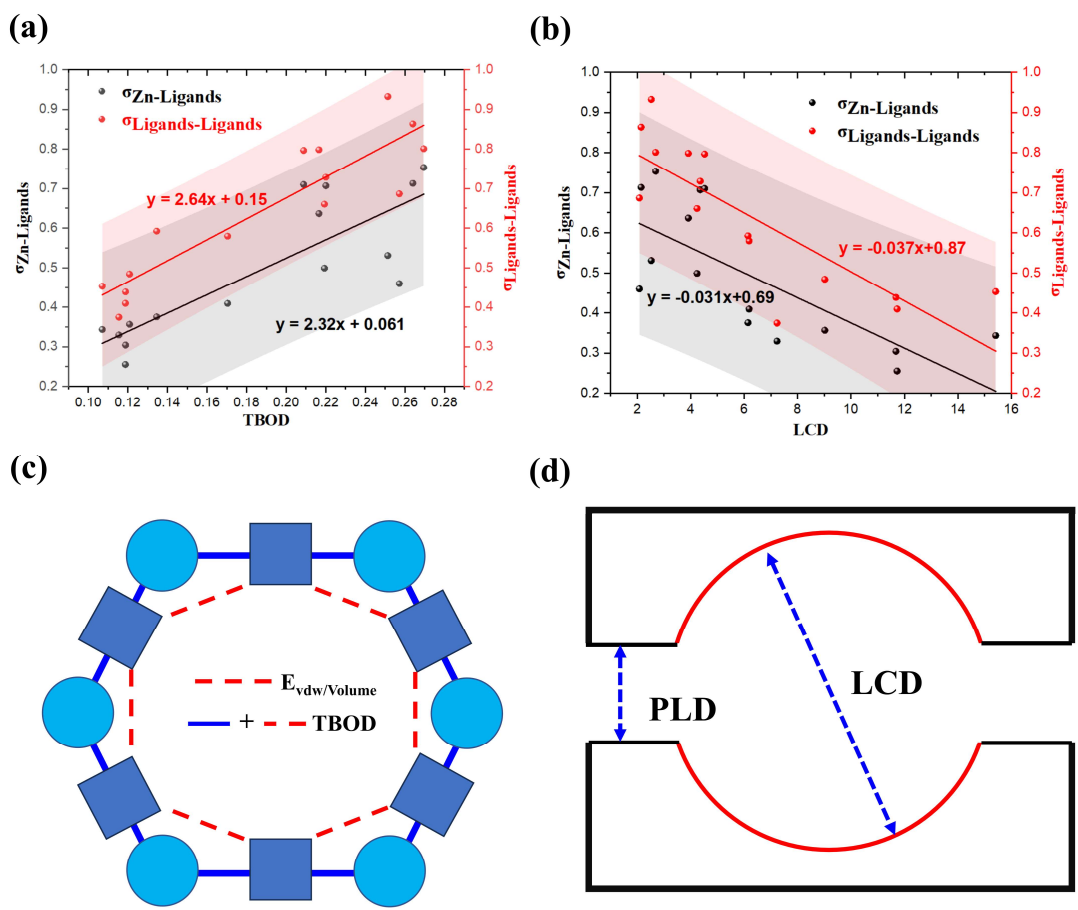

**Fig 3. Linear relationship between MOF thermal stability and effective order parameters.** (a) Changes in the radial distribution function parameters $\sigma_{Zn-Ligands}$ (black) and $\sigma_{Ligands-Ligands}$ (red) as a function of the effective order parameter TBOD and (b) LCD. (c) Schematic representation of the effective order parameters TBOD and Evdw/Volume, where the red dashed line indicates the mid-range interaction between organic functional groups and the blue solid line indicates the short-range interaction between organic functional groups and metal nodes. (d) Diagram of the effective order parameters PLD and LCD.

Meltable ZIFs necessitate the demonstration of high intra-molecular stability and low stability of metal-organic ligands, as elucidated by the microscopic analysis of ZIF melting involving short-range exchange interactions. We draw on the schematic representation of the effective order covariates in **Fig 3 (c)** and **(d)** to further explain the correlation between the melting behavior and the micro-structure of MOFs. Under appropriate temperature conditions, the detachment of organic ligands from the metal nodes occurs while ensuring their own stability, owing to the combined effect of high intra-molecular stability and low nodes-ligands stability. As demonstrated in **Fig 3 (c)**, $E_{vdw}$/Volume and TBOD indicate the strength of the MOF in terms of mid-range interaction and overall cohesion, respectively. Consequently, an intriguing speculation arises concerning ZIFs with elevated melting points due to their high thermal stability. Conversely, ZIFs with low TBOD often exhibit thermal decomposition at relatively lower temperatures. The distribution characteristics observed in the nuclear density plot partially elucidate this phenomenon, indicating the comparatively lower TBOD values for ZIFs in comparison to MOFs. Among the diverse ZIF structures, those resembling ZIF-4 exhibit glass-forming ability and possess higher TBOD values, while structures resembling ZIF-8 are more prone to thermal decomposition. At the microscopic level, the geometric arrangement of pores, as represented by LCD, exerts an influence on the thermal stability of the molecules. Referring to **Fig 3 (d)**, the PLD represents the maximum diameter of the pore channel in a porous material and may be related to the strength of the intermolecular interaction. The LCD, on the other hand, indicates the maximum endospore radius and can be considered as an "internal reaction cavity" with a domain-limiting effect. This implies that the short-range exchange process allows for the dissociation and transfer of organic ligands between nodes. Properly sized pores serve as confined reaction cavities. Structures with excessively small pores impede a significant number of organic ligands from undergoing bond-breaking and reformation processes, whereas excessively large pores result in the concentration of heat within the cavity, thereby rendering the organic ligands more susceptible to thermal decomposition.

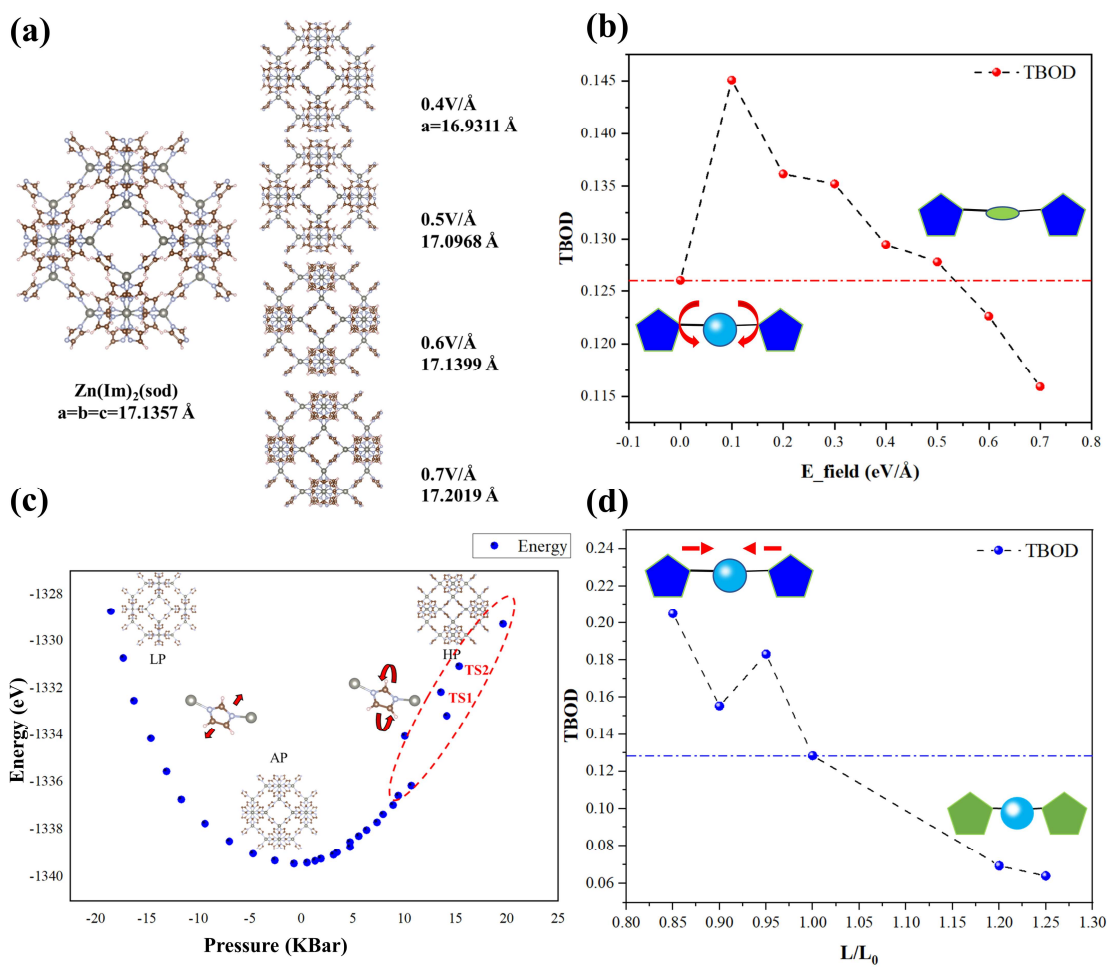

**Fig 4. Simulation of TBOD modulation of MOFs.** (a) SALEM-2 structures change under the external electric field by DFT calculations. (b) Variation of the TBOD of SALEM-2 at different electric field strengths, the inner diagram demonstrates the deflection of the functional groups due to the applied electric field. (c) SALEM-2 structures change under the external pressure by DFT calculations. (d) Variation of the TBOD of SALEM-2 corresponding to different cell covariates, the inner diagram demonstrates the short-range structural changes under external pressure.

The negative correlation between TBOD and LCD is shown in the heatmap (**Fig 1(a)**). Building upon this finding, we delve deeper into the manipulation of MOFs' glass-forming ability by modulating the TBOD. By definition, TBOD can be enhanced through an increase in the total bond order (denominator) and a decrease in the volume of MOFs (numerator). In **Fig 4** we show two ways for modulating the TBOD of SALEM-2, a ZIF exhibiting SOD topology. Analysis of the nuclear density distribution before reveals that for SOD-type ZIF-8-like ZIFs, increasing TBOD can potentially augment their glass-forming ability. For instance, the introduction of ionic liquids, construction of heterogeneous structures with more interfaces, or the application of external electric fields.[30] Recent studies have demonstrated the significant improvement of the melting stability of ZIF-8-like ZIFs with the introduction of ionic liquids.[12] Furthermore, the construction of heterogeneous structures based on high porosity MOFs is considered a feasible pathway for glass formation. [31-33] Here we explore the direct modulation of TBOD through theoretical calculations. Density functional theory calculations were performed to optimize the structure of SALEM-2 under an external electric field. **Fig 4(a)** illustrates the relationship between the lattice parameters and external electric field strength. The results reveal that the structural parameters of SALEM-2 change and lead to an increase in TBOD under an electric field intensity of 0.1 eV/Å, as depicted in **Fig 4(b)**. As the electric field further increases, a noticeable structural phase transition occurs, resulting in a reduction of TBOD due to the accompanying structural rearrangement. When the electric field intensity exceeds 0.5 eV/Å, the TBOD of SALEM-2 becomes lower than that of the ground state structure. However, it is worth noting that the electric field intensity applied in our study is exceptionally high. Considering the size effects of the atomic model (length of structure < 10 nm), it is possible to induce structural phase transitions and alter TBOD with electric field intensities much lower than 0.1 eV/Å in actual experiments.

The external electric field affects the TBOD by changing the spatial orientation of the organic ligands in SALEM-2, while another more intuitive approach is to compress the MOF by applying pressure, as shown in **Fig 4(c)**. It is obvious that the change from negative to positive pressure induces a change in the SALEM-2 lattice parameters. With the change of lattice parameters due to the external pressure, the TBOD at varied lattice parameters of SALEM-2 in **Fig 4(d)**. It can be observed that as the unit cell parameters decrease, the TBOD of SALEM-2 gradually increases. However, excessively high pressure can induce distortion of functional groups in the geometry, resulting in a decrease in TBOD, as evidenced by the comparison between the results at $L/L_0$=0.9 and 0.95. This implies that there exists a reasonable range for adjusting TBOD via pressure. Considering experimental considerations, the application of external pressure to modify lattice parameters and consequently increase TBOD presents a viable approach to enhance glass-forming ability. This approach has been reported in related studies on ZIF-4-like materials.[9,34] Despite the absence of documented experiments specifically investigating ZIF-8-like structures, given the observed pressure sensitivity of flexible ZIF-8,[35] we hypothesize that moderate pressure is advantageous in preventing the high-temperature decomposition of ZIF-8.

**Conclusion and Discussion**

In this work, we first compared the correlation between different order parameters of MOFs based on the QMOF database. Referring to existing experimental data, we selected two types of order parameters that could be related to the glass-forming process. One is the order parameter describing the electronic structure: total bond-order density (TBOD) and van der Waals energy ($E_{vdw}$/Volume). Both reflect the strength of the bonding interactions in MOFs. The other type is the order parameters describing the geometric structure, namely the largest cavity diameter (LCD) and pore limiting diameter (PLD), which reflect the distribution characteristics of the pores in MOFs. Furthermore, we analyzed the differences in the distribution of order parameters between a subclass of ZIFs with glass-forming ability and other MOFs. We also discussed the effectiveness of these order parameters.

Subsequently, we performed reactive molecular dynamics simulations on 16 types ZIFs, simulating their thermal behavior during heating from 300-1000 K. We used TBOD and LCD as structural descriptors and quantified the changes in the bonds within the structures during the temperature increase, establishing a simple QSAR model. We combined data mining and molecular simulations to propose that TBOD, which describes the bonding behavior, and LCD, which represents the maximum pore size, can be used to provide a simple description of the thermal stability of ZIFs structures. It is worth mentioning that these two parameters can be seen as the focus of different researchers. For example, TBOD based on charge distribution is more easily understood by theoretical computational researchers, while experimental researchers find it more physically meaningful to manipulate pore size through experimental means.

Herein, we further discussed some modification methods for effectively modifying TBOD, such as pressure, solution environment, and even electric fields. However, in practical preparations, the conditions may be more complex, and it is necessary to explore possible modification methods based on specific structural properties. For stable melt-state MOF structures, good thermal stability is a necessary but not sufficient condition. Especially in actual manufacturing, we strive to achieve the melting of MOFs at lower melting points while avoiding thermal decomposition. This requires TBOD and LCD to have a reasonable distribution of values. Therefore, we believe that different adjustment methods should be applied to different MOFs. For MOF structures that are prone to thermal decomposition, increasing TBOD can enhance their thermal stability and augment the resilience of their molten state against decomposition. On the other hand, for MOFs with high thermal stability and higher melting points, it is possible to reduce the melting point by negative pressure adjustment or electric field modulation by reducing TBOD.

Based on simulations and data mining at different scales, we propose that electronic structural order parameters, namely TBOD and $E_{vdw}$/Volume, along with geometric parameters LCD and PLD, can serve as structural genes to comprehensively describe the structural characteristics of different MOFs. The correlation between these order parameters and the melting behavior of MOFs can provide valuable guidance for the development of novel MOF glasses.

**Methods**

**Data Mining**

In contrast to prevailing MOF databases, the QMOF database distinguishes itself through the implementation of a sophisticated screening methodology, which permits the curation of an expansive assemblage encompassing more than 20,000 MOF structures. This comprehensive compilation is subsequently subject to high-throughput Density Functional Theory (DFT) calculations, thereby enhancing the database's scientific rigor and computational efficiency. Consequently, QMOF not only encompasses diverse geometric properties but also encompasses electronic structure information. Moreover, complementary to the raw data within QMOF, several structural order parameters have been evaluated by batch processing geometric structures, including metal-organic node bond lengths and bond angles. All the pertinent structural parameters and their corresponding descriptions utilized in this research are cataloged in **Table S1**. It is pertinent to acknowledge that within the QMOF database, an ensemble of order parameters is encompassed, comprising the fundamental order parameters denoted as basic order parameters. Moreover, an additional subset of order parameters comprises composite order parameters derived through a synergistic amalgamation of the aforementioned basic order parameters. The energy per unit cell volume, denoted as E/Volume, is an indicator of the energy level for different MOFs, and is employed to compensate for the significant variations in cell sizes across diverse MOFs. To further characterize the chemical bond distribution features of complex structures, the bond order density (TBOD) is calculated as the total bond order divided by the cell volume, expressed as, $TBOD = \frac{\sum BO}{Volume}$. Here, the bond order (BO) is determined to use the DDEC6 charge method, which effectively represents the charge distribution of atoms in periodic structures. Notably, TBOD has found extensive utility in the study of inorganic glasses, effectively capturing the intricate chemical bonding arrangements within such structures.

**Simulation**

In this work, we conducted molecular dynamic simulations to investigate the thermal behavior of sixteen distinct zeolitic imidazolate frameworks (ZIFs) and characterized the resulting structural changes as a function of temperature. In order to strike a balance between computational efficiency and accuracy, we employed reactive force field molecular dynamics (ReaxFF) simulations.[36] Throughout the heating process, the simulations were performed using the NPT ensemble, where pressure effects were neglected to mimic volume changes induced by temperature variations. The temperature range considered spanned from 300 K to 1000 K, with a heating rate of 1 K/ps. To ensure the dynamic stability of the structures, an initial relaxation step was conducted for a duration exceeding 100 ps at 300 K. Each simulation encompassed a time period exceeding 1 ns, inclusive of the pre-equilibration heating stage. To minimize statistical uncertainties, all structural analyses were averaged over twenty snapshots captured at different temperatures.

In addition to molecular dynamic simulations, we have used density function theory (DFT) calculations in this work to investigate the electronic structure of MOFs. All DFT calculations were performed using the Vienna Ab initio Simulation Package (VASP) software.[37] All geometric optimizations and electronic structure calculations were carried out employing the Generalized Gradient Approximation with Perdew-Burke-Ernzerhof (GGA-PBE) functional within the Projector Augmented Wave (PAW) method.[38,39] The DFT-D3 method was employed to address the shortcomings of the chosen functional in describing long-range interactions in MOFs.[23] A 3x3x3 Monkhorst-Pack k-point mesh was used for sampling the Brillouin zone. The energy cutoff was set to 500 eV, and the convergence criteria for electronic and ionic steps were set to $10^{-5}$ eV and -0.001 eV*Å$^{-1}$, respectively. The DDEC6 charge scheme was employed for handling the charge distribution,[40,41] followed by post-processing using the Chargemol program.[42]

**Acknowledgements**

This work was supported by the Opening Project of State Key Laboratory of Advanced Technology of Float Glass (No. 2022KF02), the Fund of National Natural Science Foundation of China (No. 11604249), the Fok Ying-Tong Education Foundation for Young Teachers in the Higher Education Institutions of China (No. 161008), the Opening Project of State Key Laboratory of Refractories and Metallurgy, Wuhan University of Science and Technology (No. G201605), and the Fundamental Research Funds for the Central Universities (No. WUT35401053-2022).